\documentclass[final]{svjour2}
\usepackage{graphicx}
\usepackage{rotating}
\usepackage{amssymb}
\usepackage{mathptmx}
\usepackage[numbers]{natbib}
\makeatletter
\journalname{Journal of Low Temperature Physics}

\bibpunct{}{}{,}{s}{}{,}

\begin{document}

\newcommand{\hdblarrow}{H\makebox[0.9ex][l]{$\downdownarrows$}-}
\title{Cosmology Large Angular Scale Surveyor (CLASS) Focal Plane Development}

\author{D.T. Chuss \and A. Ali \and M. Amiri \and J. Appel \and C.L. Bennett \and F. Colazo \and K.L. Denis \and R. D\"{u}nner \and T. Essinger-Hileman \and J. Eimer \and P. Fluxa \and D. Gothe \and M. Halpern \and K. Harrington \and G. Hilton \and G. Hinshaw \and J. Hubmayr \and J. Iuliano \and T.A. Marriage \and N. Miller \and S.H. Moseley \and G. Mumby \and M. Petroff \and C. Reintsema \and K. Rostem \and K. U-Yen \and D. Watts \and  E. Wagner  \and E.J. Wollack \and Z. Xu \and L. Zeng}

\institute{Department of Physics, Villanova University\\ Villanova, PA 19085, USA\\ Tel.:610-519-4884\\ Fax:\\
\email{david.chuss@villanova.edu}}
\maketitle

\begin{abstract}

The Cosmology Large Angular Scale Surveyor (CLASS) will measure the polarization of the Cosmic Microwave Background to search for and characterize the polarized signature of inflation.  CLASS will operate from the Atacama Desert and observe $\sim$70\% of the sky. A variable-delay polarization modulator (VPM) modulates the polarization at $\sim$10 Hz to suppress the 1/f noise of the atmosphere and enable the measurement of the large angular scale polarization modes. The measurement of the inflationary signal across angular scales that span both the recombination and reionization features allows a test of the predicted shape of the polarized angular power spectra in addition to a measurement of the energy scale of inflation.  

CLASS is an array of telescopes covering frequencies of 38, 93, 148, and 217 GHz. These frequencies straddle the foreground minimum and thus allow the extraction of foregrounds from the primordial signal. Each focal plane contains feedhorn-coupled transition-edge sensors that simultaneously detect two orthogonal linear polarizations. The use of single-crystal silicon as the dielectric for the on-chip transmission lines enables both high efficiency and uniformity in fabrication.  Integrated band definition has been implemented that both controls the bandpass of the single mode transmission on the chip and prevents stray light from coupling to the detectors. 


\end{abstract}

\section{Introduction}
Measurement of the polarization of the cosmic microwave background (CMB) provides a means to test and constrain inflation.  Inflation provides a mechanism for producing the observed flatness of the universe, near uniformity of the CMB, and the nearly scale invariant scalar perturbations in the density of the early universe.  In addition to these density perturbations, inflation predicts a spectrum of gravitational waves that will polarize the CMB.   Whereas polarization that arises from the density fluctuations will contain only a curl-free component (``E-modes''), the inflationary gravitational waves will contain both E-modes and a divergence free set of ``B-modes.'' Thus, a search for CMB B-mode polarization provides a well-posed test of inflationary physics.  In addition to testing inflation, CLASS's measurement of the E-mode polarization to unprecedented  precision will provide significant improvements in the measurement of the optical depth to reionization. These measurements will distinguish between reionization scenarios and, in conjunction with data from higher resolution instruments, will help constrain neutrino masses. 

The challenges in measuring the CMB polarization in search of inflation stem from the small expected amplitude of the signal and the fact that it needs to be measured in the presence of an unpolarized signal that is 8-9 orders of magnitude larger than the desired signal.  These two conditions push instrument designs to minimize systematic errors, but also maximize sensitivity.  The latter criterion motivates focal planes that have sufficient sensitivity to make the measurement with feasible integrations times.  There are two features that are important here. First, the detectors should be sensitive enough such that photon noise dominates over detector noise.  For single-mode sensors, throughput is proportional to the number of sensors in the focal plane; however, the number of detectors does not by itself constitute an effective metric for sensitivity. The efficiency of the sensor architecture along with that of the instrument are also important in determining array sensitivity. In this work, we describe the CLASS focal plane technology, focusing on the completed 38 GHz array. 

\section{CLASS}
The Cosmology Large Angular Scale Surveyor (CLASS) is a telescope array designed to measure the CMB at a unique combination of angular scales and frequencies. A more complete summary of the instrument is given in Essinger-Hileman et al. (2014),\cite{Essinger-Hileman14} but the main features will be summarized here. The four focal planes consist of the following. One focal plane with 36 detector pairs operating at 38 GHz, two focal planes with 259 detector pairs each operating at 93 GHz, and a dual-band telescope with $\sim$1000 detector pairs at each of 148 and 217 GHz. The NEQ per single polarization mode is expected to be 208 $\mu K\sqrt{s}$ for the 38 GHz receiver that is described here. The combined efficiency of the optics and the detector is 68\%. 

CLASS is located at an altitude of $\sim$5200 feet in the Atacama Desert. The CLASS survey will cover 70\% of the sky in order to access multipoles ($\ell<10$) associated with the reionization feature of the predicted B-mode spectrum. The large-scale survey is enabled by front-end polarization modulation provided by a variable-delay polarization modulator \cite{Chuss06, Chuss12} for each of the four telescopes. Rapid ($\sim 10$ Hz) polarization switching is used to suppress the long term ($1/f$) fluctuations of the atmospheric emission.  Secondary modulations such as constant elevation scanning and periodic rotations about the boresight will be used to decouple instrumental effects and fill in the parameter space for the survey.

CLASS will have sensitivity below a tensor-to-scalar ratio, $r=0.01$.  Techniques foreground removal using the 4 CLASS bands\cite{Watts} and mitigation of systematic effects associated with using a VPM to recover the large scale polarization modes\cite{Miller} have been validated through simulations and have been found to be effective for the target $r$.

\section{Detector Concept}

\begin{figure}
\begin{center}
\includegraphics[%
  width=0.6\linewidth,
  keepaspectratio]{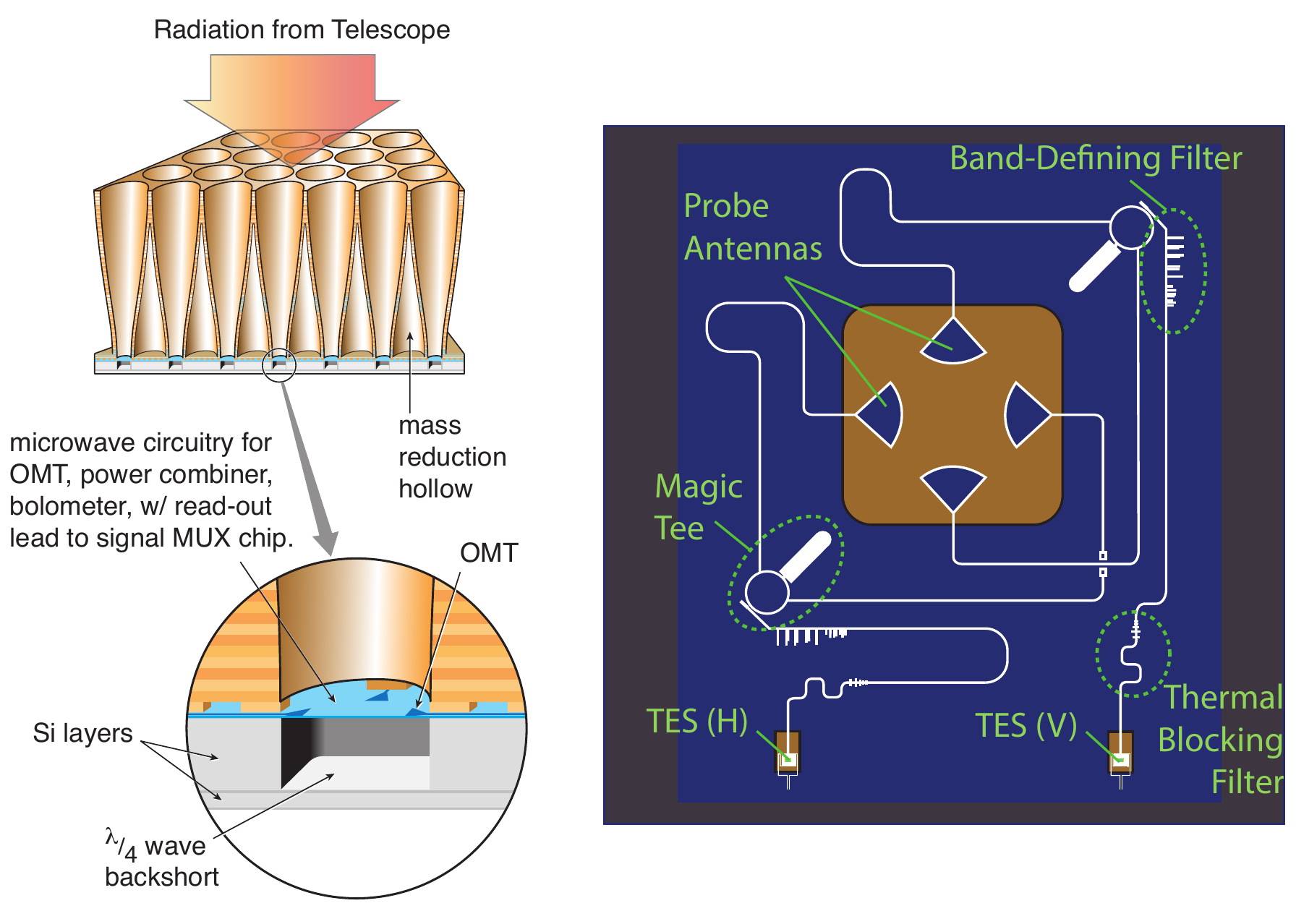}
\end{center}
\caption{ {\it Left} The schematic of the detector concept is shown. {\it Right} The layout of the 40 GHz detector chip is shown. (Color figure online.)}
\label{fig:fig1}
\end{figure}

The detector concept is shown in Figure~\ref{fig:fig1}. \cite{Stevenson08, Chuss12, Rostem14} Light from the telescope is incident on an array of feedhorns. To make fabrication less expensive, and thus improve the scalability of these arrays, the array is implemented using monotonic smooth-walled feedhorns\cite{Zeng10} that exhibit good beam symmetry and a measured cross-polarization response of better than -30 dB across each of the 38 and 93 GHz bands out to the edge of the pupil stop.  

At the base of each feedhorn is a planar orthomode transducer (OMT) fabricated in niobium microcircuitry with a monocrystalline silicon dielectric. The monocrystalline silicon provides several advantages.  First, the dielectric loss in the silicon is negligible, resulting in high efficiency for the sensor.  Second, the electrical permittivity of the material is highly uniform, reducing the variability of the electrical properties of the circuit between detectors. Third, the acoustic properties of the material are uniform, allowing precision control over thermal conductance of the devices. 

The interface between the metallic feedhorn and the silicon wafer is controlled by implementing a photonic choke,\cite{Wollack10} which consists of an array of square pillars that prevent leakage at this interface without a conductive connection. The photonic choke is implemented by direct machining the metallic part of the interface in the 38 GHz array; in the 93 GHz array, the choke is etched into the top layer of the wafer assembly. 

The OMT independently couples two orthogonal polarization modes into independent microstrip circuits. Unwanted modes are rejected by the OMT design, which consists of a magic-tee that is used to difference the signal from opposite antennas.\cite{U-Yen07, U-Yen08a} The desired (symmetric) signal for each antenna pair is coupled to the microstrip line for that polarization. The asymmetric modes are resistively terminated at the sum port of the magic-tee. 

Once the signal for each polarization is in the microstrip, a series of band-defining and thermal blocking filters limit the response of the sensor to the desired passband. For the case of the 38 GHz focal plane, this is 33-43 GHz.  For the 93 GHz focal planes, the bandpass is 77-108 GHz. 

\begin{figure}
\begin{center}
\includegraphics[%
  width=0.4\linewidth,
  keepaspectratio]{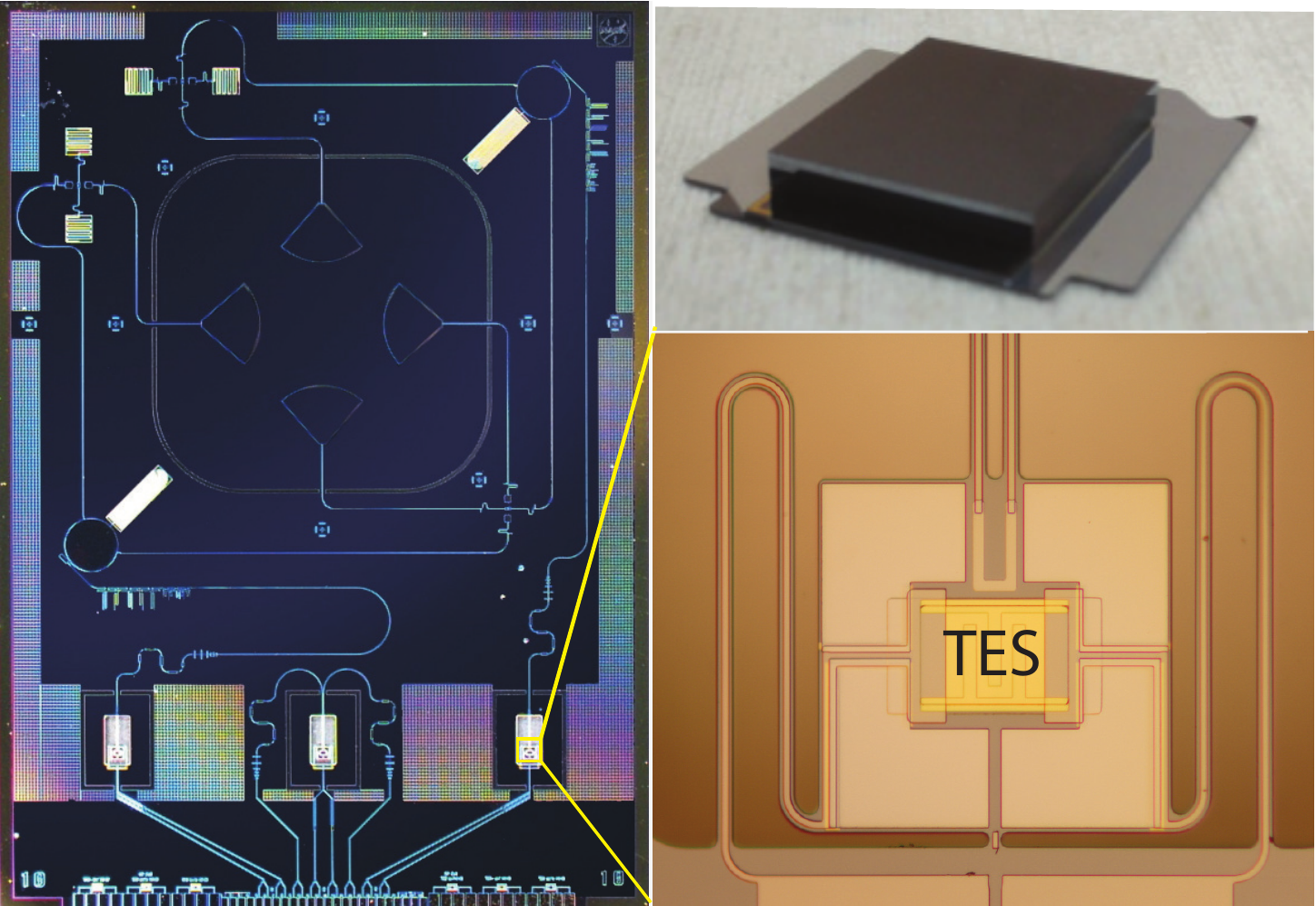}
\end{center}
\caption{ {\it Left} The 38 GHz detector chip is shown. {\it Upper Right} The fully integrated single 38 GHz dual-polarization sensor is shown.  The black box is the backshort cap that completes the OMT by maximizing the electric fields in the vicinity of the antennas as well as blocks stray light from coupling to the TESes.  {\it Lower Right} A photo of one of the TESes is shown. Thermal conductance of the detector is set by a ballistic leg connecting the TES island to the substrate, visible at the bottom of the TES island (Color figure online.)}
\label{fig:circuit}
\end{figure}

After filtering, the signal is terminated in a resistor that is thermally coupled to a transition-edge sensor (TES) that has a target transition temperature of $T_c$=150 mK. The circuit is shown in the left panel of Figure~\ref{fig:circuit}. The TES is shown in the lower right panel of the same figure. The microwaves are transmitted to the absorber by the two straight legs near the top of the photo. The curved legs on either side are
the TES bias leads. These long legs contribute very little to the thermal conductance of the TES to the frame because their side walls are rough. The thermal conductance is primarily set by the ballistic phonon conduction in the short leg near the bottom of the image.\cite{Rostem14a}  

In the interest of minimizing systematic errors, stray light control is essential.  As mentioned above, out-of-band radiation that enters the detector via the microstrip lines is blocked by the thermal blocking filters (and by the waveguide cutoff in the feedhorn on the low frequency side).  Direct stray light coupling to the TES is minimized by incorporating radiation shielding into the backshort cap (see Fig~\ref{fig:circuit}, top right). Fabrication has been described in previous work.\cite{Denis09, Crowe14, Denis15}

\begin{figure}
\begin{center}
\includegraphics[%
  width=0.45\linewidth,
  keepaspectratio]{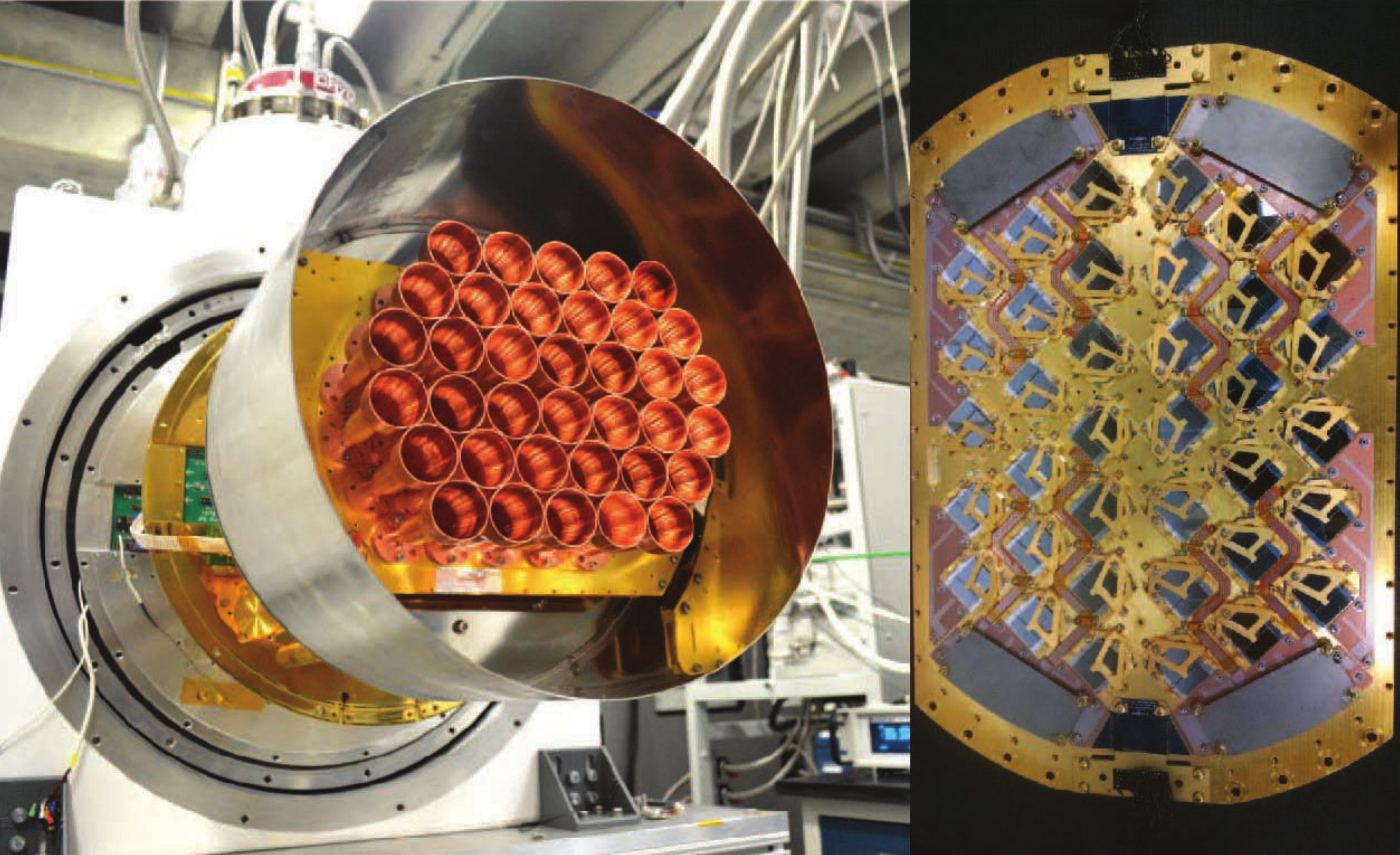}
\end{center}
\caption{ {\it Left} The CLASS 38 GHz focal plane is shown from the feedhorn side. {\it Right} The back of the focal plane consists of 36 dual polarization individual detectors that are accurately mounted to a single plate. (Color figure online.)}
\label{fig:FP}
\end{figure}

\section{CLASS Focal Planes}
Each CLASS focal plane is housed in a separate cryostat having lenses that are cooled to 4 K and 1 K via a pulsed tube cooler. The focal plane is cooled to 100 mK via a continuous dilution refrigerator. A 4 K stop is located near the window of the cryostat. The pupil is re-imaged by warm reflective optics onto the 60 cm diameter VPM, which is located at the front of the optics.\cite{Eimer12}

The CLASS 38 GHz focal plane is complete and shown in Figure~\ref{fig:FP}.  The focal plane consists of 36 detector pairs, each containing a feedhorn-coupled TES for each polarization. Because of the large size required at this frequency, each detector pair is individually fabricated, integrated with its backshort cap, and mounted to the focal plane.  This process has resulted in 98\% of the TESes operational.  The detector noise at the modulation frequency has been measured to be 10-15 $aW Hz^{-1/2}$ for all sensors. The measured efficiency for the detector is $\sim90\%$, and using a waveguide calibration source,\cite{Rostem13} we have verified no appreciable out-of-band leaks up to $\sim500$ GHz. 

\begin{figure}
\begin{center}
\includegraphics[%
  width=0.75\linewidth,
  keepaspectratio]{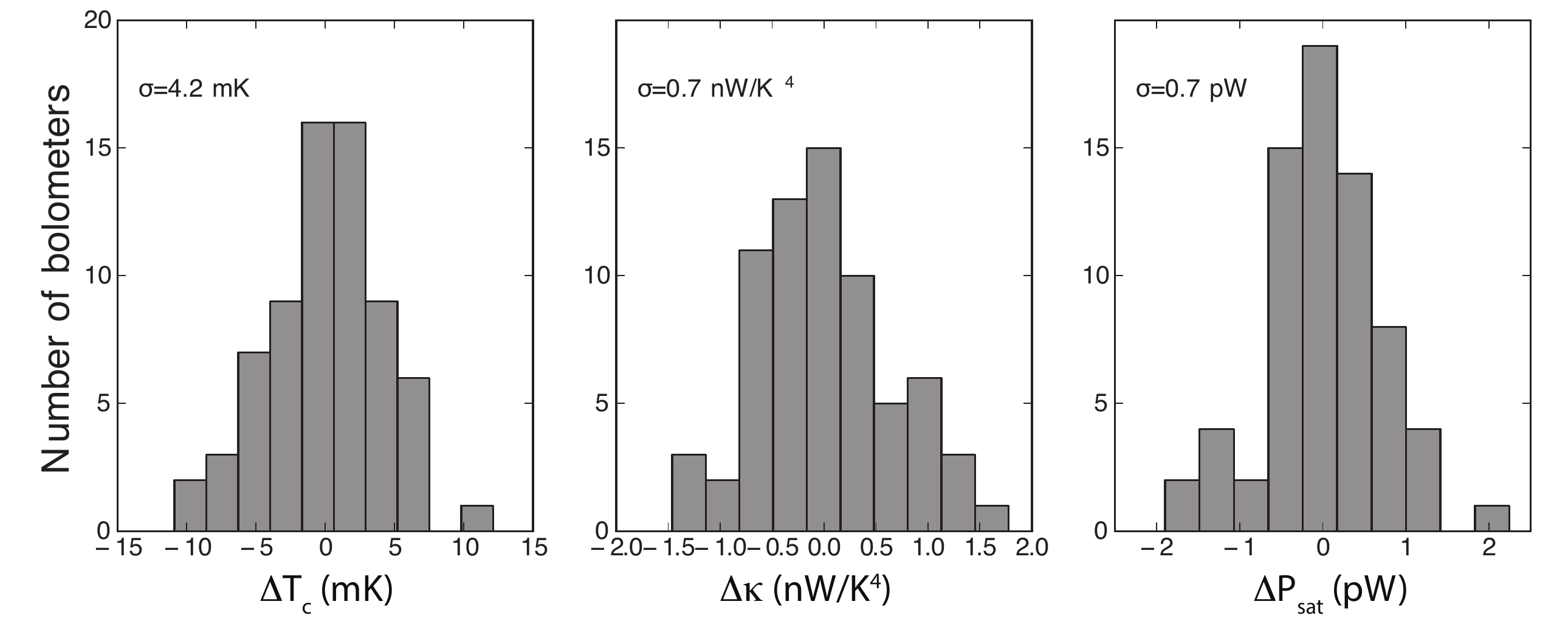}
\end{center}
\caption{The intra-wafer spread of the critical temperature ($T_c$), thermal conuctance, $\kappa$, and saturation power ($P_{sat}$) are shown for the 38 GHz focal plane.}
\label{fig:hist}
\end{figure}

Histograms of the deviation from the wafer average for achieved $T_c$, thermal conductivity ($\kappa$), and saturation power ($P_{sat}$) are shown in Figure~\ref{fig:hist}. These are similar to those previously reported,\cite{Appel14} but have been updated to include the entire 38 GHz CLASS focal plane.

\begin{figure}
\begin{center}
\includegraphics[%
  width=0.65\linewidth,keepaspectratio]{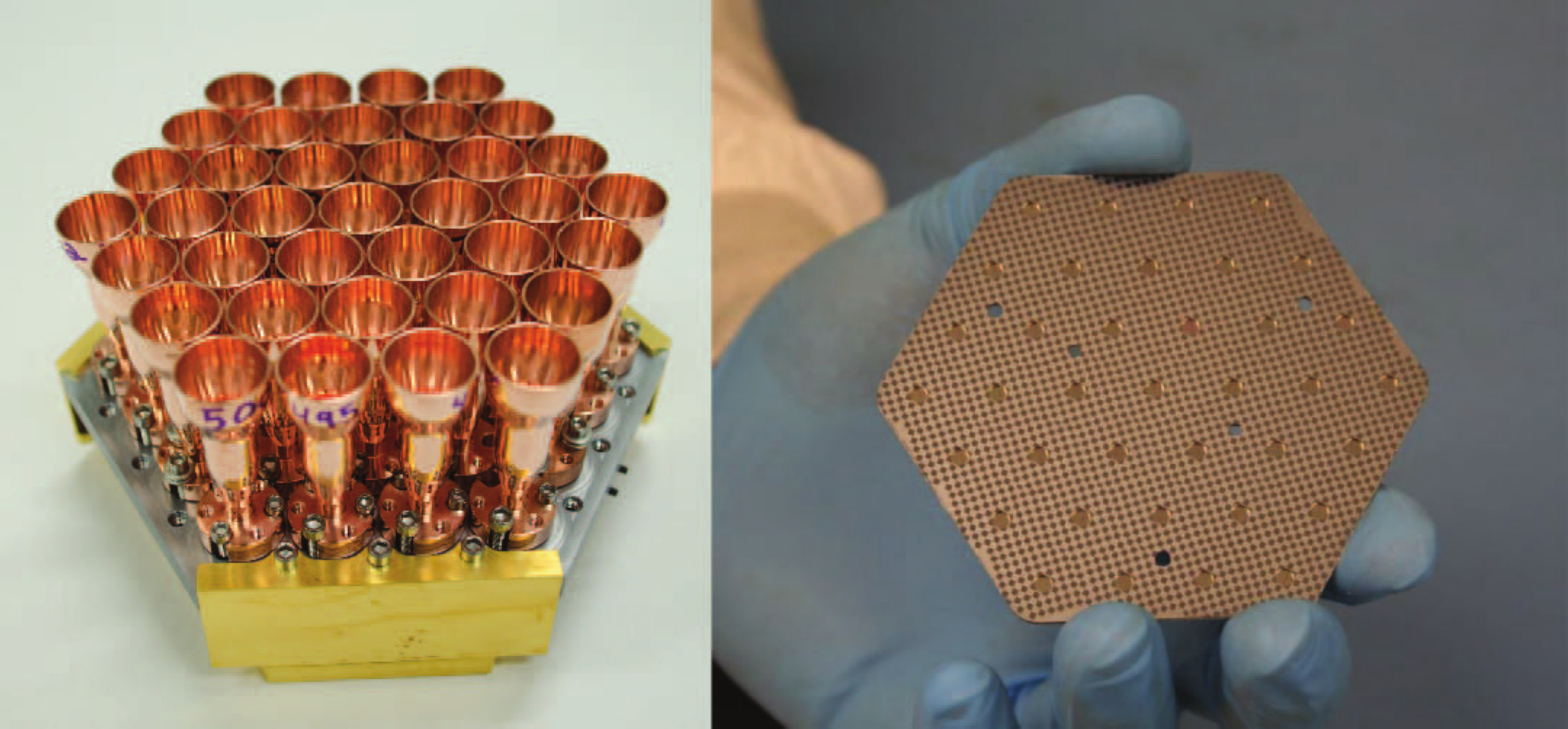}
\end{center}
\caption{ {\it Left} A CLASS 93 GHz detector module is shown.  {\it Right} A fully integrated 37-element CLASS 93 GHz wafer is shown. The pattern of squares visible on the surface are the photonic choke pillars (Color figure online.)}
\label{fig:wband}
\end{figure}

The 38 GHz focal plane is currently integrated into the CLASS cryostat and will be deployed to the Atacama in late 2015.  Additional receivers will follow. For frequencies higher than 38 GHz, focal planes consist of tilable modules. Photographs of a 93 GHz module and a 93 GHz integrated wafer are shown in Figure~\ref{fig:wband}.

\begin{acknowledgements}
Support for CLASS has been provided by the National Science Foundation through grant numbers 0959349 and 1429236. The NASA ROSES/APRA program has provided funding for the development of the detectors. K. Harrington was supported by a NASA Space Technology Research Fellowship (NNX14AM49H). D Watts is funded by the Maryland Space Grant Consortium.
\end{acknowledgements}


\end{document}